\documentstyle[epsf,12pt]{article}

\begin{document}
\setlength{\baselineskip}{0.30in}

\begin{flushright}
UM - TH - 97 - 04\\
\today\\
hep-ph/9702270
\end{flushright}

\begin{center}
\vglue .06in
{\Large \bf{ The KLN Theorem and Soft Radiation in Gauge Theories:
Abelian Case}}\\[.5in]

{\bf R. Akhoury, M. G. Sotiropoulos and V. I. Zakharov}\\[.15in]

{\it{The Randall Laboratory of Physics\\
 University of Michigan\\
 Ann Arbor, MI 48109-1120}}\\[.15in]

\end{center}

\begin{abstract}
\begin{quotation}
 We present a covariant formulation of the  Kinoshita, Lee, Nauenberg (KLN)
 theorem  for processes involving the radiation of soft particles.
 The role of the disconnected diagrams is explored  and 
 a rearrangement of the perturbation theory is performed such that the purely
 disconnected diagrams are factored out.  
 The remaining effect of the disconnected diagrams results in a 
 simple modification of the usual Feynman rules for the S-matrix elements.
 As an application, we show that when combined with the Low theorem, 
 this leads to a proof of the absense of the $1/Q$ corrections to 
 inclusive processes (like the Drell-Yan process).  
 In this paper  the abelian case is discussed to all orders in the
 coupling.
\end{quotation}
\end{abstract}

\newpage

\section{Introduction }

 The presence of massless particles in gauge theories brings about the
 problem of  infrared divergences.  
 In particular, it is well known that the cross section of, say, 
 single photon emission is infrared divergent in the limit  of vanishing  
 photon energy. 
 The way out of the difficulty was first found by Bloch and Nordsieck 
 \cite{bn}, who showed that the infrared divergence cancels out provided that
 we consider inclusive processes in which the bremsstrahlung contribution is
 combined with the radiative corrections to the elastic process. 
 In this way a finite resolution $\Delta\neq 0$ in the photon energy is 
 introduced.

 The  most general framework for these kind of cancellations is provided  
 by a quantum-mechanical theorem, due to Kinoshita,  Lee and Nauenberg (KLN) 
 \cite{kln}. 
 According to this theorem all  the singularities are cancelled provided 
 that summation over all degenerate in energy states is performed. 
 An important point is that summation over both initial and final
 degenerate states is required,
\begin{equation}
\sum_{i,f}|S_{i\rightarrow f} |^2~\sim~
{\rm free~of~all~infrared~singularities} \, ,
\end{equation}
 where $S_{i\rightarrow f}$ are the $S$-matrix elements.
 Since the summation over the initial state does not correspond to 
 an experimental resolution, infrared singularities persist, 
 generally speaking, in physical cross sections. 
 From this point of view the Bloch-Nordsieck cancellation, upon the summation 
 over final states alone, looks rather as an exception than a rule. 
 The reason for this exception is that in the limit of vanishing 
 photon energy the emission and absorption of photons turn out to be 
 indistinguishable.

 With the advent of QCD the problem of infrared singularities became even
 more  acute. 
 Indeed, because of the presence of (nearly) massless quarks,
 collinear singularities are present in addition to the soft ones.
 The KLN theorem is instrumental in isolating the collinear singularities  
 in physical cross sections such that the effect of these 
 singularities is embedded into phenomenological parton distribution functions
 $f_{a/h}(x)$, where $x$ is the momentum fraction of the parent hadron $h$
 carried  by parton $a$.  
 As an example of an inclusive process we will consider the cross section 
 of  the Drell-Yan (DY) process, $h_1+h_2\rightarrow \mu^+\mu^- + X$ . 
 The cross section is then  given by 
\begin{eqnarray}
{ {\rm d} \sigma\over {\rm d} Q^2}(\tau,Q^2)~=~\sum_{a,b} 
\int_0^1 {\rm d} x_a \int_0^1 {\rm d} x_b \int_0^1 {\rm d} x \, 
\delta(\tau-x_a x_b x) \,
 f_{a/h_1}(x_a) \, f_{b/h_2}(x_b) \,
\left[ \sigma_0 W_{ab}(x,Q^2) \right] , 
\label{dy} \\
\sigma_0 = {4\pi \alpha_{QED}^2 \over {9 Q^4}} \tau \, ,
\end{eqnarray}
 where $s, Q^2$ are squares of the hadronic and leptonic invariant masses 
 respectively, $\tau=Q^2/s$ and $\left[ \sigma_0 W_{ab}(x,Q^2) \right]$ 
 is the hard partonic inclusive cross section for 
 $a+b \rightarrow l^+l^- + {\rm gluons}$. 
 In the following we restrict our attention to the case when
 the partons are the quarks and antiquarks, 
 $\{a,b\} = \{q, \bar{q} \}$.
 Soft gluon radiation from these can be the origin of possible
 $1/Q$ corrections, which will be the main application of the
 general statements presented in this paper.
 The hard cross section is perturbatively calculable
 and the parton distribution functions can be deduced from some other
 process, such as deeply inelastic scattering.

 Another way to handle the problem of infrared divergences is through
 the introduction of infrared safe quantities. 
 A typical example of this kind, which we  will have in mind, is the thrust, 
 $T$, in $e^+e^-$  annihilation,
\begin{equation}
T~=~{\rm max}_{{\bf n}}
{\Sigma ({\bf p}_i{\bf n})\over \Sigma |{\bf p}_i |} \, ,
\label{thrust} 
\end{equation}
 where ${\bf p}_i$ are the momenta of particles while ${\bf n}$ is 
 a unit vector.
 The leading contributions to these observables are calculable 
 in perturbation theory and depend logarithmically on the total energy.

 Nonperturbative effects induce power like corrections to infrared safe
 quantities. 
 The theory of renormalons (for a review see Ref. \cite{az4}) is a 
 newly emerging 
 method which provides an indication of the form of these  corrections.
 The presence of nonperturbative corrections is signaled by divergences of
 the perturbative expansions in large orders. 
 The asymptotic nature of the perturbative expansion implies that
 the series cannot approximate a physical quantity beyond a certain
 accuracy. 
 The uncertainties revealed by the renormalons are power like, i.e.
 they are proportional to $1/{Q^p}$, where $Q$ is the large momentum scale. 
 
 Of particular interest are the leading power corrections which turn out 
 to be linear  in $1/Q$, $\sim \Lambda_{QCD}/Q$  \cite{cs}-\cite{bb}.
 Such terms do not jeopardize the calculability of various observables but 
 provide us with a measure of their infrared sensitivity. 
 These leading power corrections have important phenomenological relevance  
 in particular  for the extraction of the strong coupling $\alpha_s$.  
 Theoretical investigations indicate 
 that  they arise due to soft gluon radiation hence providing us with 
 another interesting consequence of the latter.

 In a previous letter \cite{az3} arguments were presented that linear terms 
 and logarathmic divergences share not only the property of universality but a
 Bloch-Nordsieck type of cancellation as well. 
 Namely, if one considers inclusive cross sections, that is a case of poor 
 energy resolution, the linear terms cancel. 
 If, on the other hand, the accuracy of the measurements on the final state is 
 of the order of an infrared parameter, then linear terms survive. 
 As an example of an inclusive process we will consider the Drell-Yan 
 cross section. Observables which assume precision measurements are 
 exemplified by thrust.
 This distinction between inclusive and exclusive processes provides a 
 general framework to interpret the results of explicit one-loop
 calculations \cite{bb}.

 It is worth emphasizing that this extension of Bloch-Nordsieck type of
 cancellation becomes possible only because the $1/Q$ corrections are
 determined by soft particle emission. 
 That is why in the KLN sum, which eliminates $1/Q$ terms \cite{asz},  
 the summation over 
 degenerate initial states is not separated by a large energy gap from the
 summation over final states, which corresponds to the evaluation of an
 inclusive cross section.
 Furthermore, the bridge between the two summations  to the needed accuracy
 is provided by the Low theorem \cite{low}, which provides a universal recipe
 to evaluate linear corrections to the amplitude of soft particles.
 In this respect the situation is very different from the case of collinear
 divergences when the KLN summation includes energetic particles in the
 initial state and cannot be related to any inclusive cross section.

 The purpose of the present paper is,  apart from providing  all the relevant 
 technical details,  to extend the analysis of \cite{az3} to all orders. 
 In the course of this we have developed a novel covariant formulation 
 of the  KLN theorem for soft radiation which we hope will be useful for 
 other applications. 
 It has been argued  \cite{asz} that the sum over both initial and final 
 degenerate states envisaged in the  KLN theorem removes not only
 the terms logarithmic in an infrared cutoff but those linearly
 dependent on it as well. 
 However, since the KLN sum involves summing the square of the S-matrix 
 over both initial and final states we need to do some more work to extract 
 from this a statement about physical inclusive cross sections. 
 We show how this is possible by combining the Low theorem 
 with the KLN theorem.
 The application of the Low theorem in the high energy regime entails
 new considerations.
 In particular, once one includes radiative corrections to soft emission,
 the collinear singularities get mixed up at the intermediate stages 
 and apparently put limitations on the possibility of using
 an expansion in the energy of the soft particles.
 This kind of problems has been discussed in Ref.~\cite{dd} and we will
 return to it in a later section.
 In this paper we will consider the case of abelian gauge bosons in detail 
 and take up the non-abelian case in a future publication \cite{asz1}.

 A precise formulation of the KLN theorem can be made in terms of the quantity
\begin{equation}
{\cal P}_{mn}~=~{1\over m!}{1\over n!}\sum_{i,f}|M_{mn}|^2 \, ,
\label{mn}
\end{equation}
 where $M_{mn}$ is the amplitude for a general radiation process
\begin{center} 
 $ A + m \ \mbox{soft~photons}\ \rightarrow 
   B + n \ \mbox{soft~photons}\ $. 
\end{center}
 In general, ${\cal P}_{mn}$ contains contributions from 
 disconnected diagrams of $M_{mn}$. 
 We will refer to ${\cal P}_{mn}$ as the Lee-Nauenberg (LN) probabilities.
 The assertion of the KLN theorem is that the quantity ${\cal P}$,
\begin{equation} 
{\cal P}~=~\sum_{m,n}{\cal P}_{mn}
\end{equation}
 does not contain either terms logarithmically \cite{kln} or linearly 
 \cite{asz} dependent on an infrared cutoff $\lambda$ 
 (which apart from other possibilities can be $(k_{\perp})_{min}$, 
 or the mass of a U(1) gauge boson).
 The contribution of the disconnected diagrams ensure that even for
 the simplest case of single gauge boson emission an infinite
 number of diagrams contribute to the LN probabilities. 
 However, we will show that by a suitable rearrangement of the KLN sum 
 the contribution of the purely disconnected diagrams can be factorized out 
 from the connected diagrams that contribute to the probabilities. 
 The latter are finite in number for an amplitude
 involving a finite number of photons. 
 We call this procedure KLN factorization and in section 2 we deal with
 the case of one photon emission or absorption in the context
 of the Drell-Yan process. 
 Thus for this example, $A$ is $q + \bar{q}$ and $B$ is $\gamma^*$, 
 the off-shell photon that decays into the lepton pair. 
 This example sets the stage for a general all orders treatment which 
 we discuss in section 3.
 We would like to emphasize that this treatment of soft radiation in the 
 context of the KLN theorem is novel and interesting in its own right. 
 We have reformulated the KLN theorem in covariant language and moreover 
 in a manner which allows us to deal with the disconnected
 diagrams in a simple and consistent way. 
 In particular, we find an interesting rule.
 This is that the effect of the
 disconnected diagrams is essentially to replace the usual gauge
 propagators in the diagrams for the inclusive cross section by itself and
 its complex conjugate. 
 This reformulation may find applications outside of the area of power 
 corrections though the latter was the motivation for this study.
 In section 4 we discuss the Low theorem \cite{low} for multiple soft gauge
 boson processes following the work in Ref.~\cite{dd}. 
 In section 5 we put together all our results to conclude on the absence 
 of any terms linear in an infrared cutoff in inclusive processes like the 
 Drell-Yan process. 
 In section 6 we summarize our conclusions.

\section{ A Reformulation of the KLN theorem for Processes Involving a Single
          Gauge Boson }

 Let us begin by studying a single photon process for which the
 radiationless amplitude is $q + \bar{q} \rightarrow \gamma^*$.
 Typical contributions will include the virtual diagrams like in Fig.~1. 
 Here, the small circle denotes the emission of the virtual photon $\gamma^*$.
 In fact, we can include hard interactions as well as the virtual gauge 
 particles into a single blob and we will follow this notation throughout 
 the section. 
 Thus, the diagram in Fig.~2 will be denoted by $P_{00}$ and includes 
 the contributions of the virtual photons to the relevant order inside 
 the shaded region. 
 The vertical dotted line denotes the unitarity cut.

 We can have, in addition, the diagrams with a soft photon 
 in the final state, Fig.~4 
 (arrows on the photon lines denote direction of momentum flow),
 which arise upon squaring the amplitude of the photon emission, Fig.~3.
 The set of all such cut diagrams will be denoted by $P_{01}$. 
 Finally, according to the KLN prescription for degenerate states, we must 
 consider diagrams where a soft photon is absorbed in the initial state. 
 Thus  we are led to consider in ${\cal P}_{mn}$ of Eq.~(\ref{mn}) absorption 
 amplitudes squared which correspond to the diagrams in Fig.~5.
 We denote the set of all such diagrams by $P_{10}$. 
 Note that superficially, $P_{10}$ and $P_{01}$ look similar, however, 
 the energy momentum  conservation constraints are different for the two sets. 

 In addition to the above we also have the contribution of the disconnected 
 diagrams. 
 From the interference between the connected diagrams of Fig. 6(a) and the 
 disconnected ones of Fig. 6(b) we can get contributions to ${\cal P}_{mn}$ 
 from sets of the type in Fig.~7. 
 These sets will be denoted generically by $P_{11}^{(1)}$. 
 Notice that the quark and antiquark propagators are the same in 
 $P_{00}$ as in $P_{11}^{(1)}$. 
 In fact it is easy to check that $P_{11}^{(1)}$ can be obtained 
 from $P_{00}$  by replacement of the virtual photon line by
\begin{equation}
{i\over k^2+i\epsilon}~\rightarrow~ 2\pi\delta(k^2).
\label{repl}
\end{equation}
 The contribution from disconnected diagrams do not end here. 
 In fact, an infinite number of disconnected photon lines can be added 
 without changing the order of perturbation theory.
 Consider for example the diagrams in Fig.~3. 
 If we add a disconnected photon line (see Fig.~8)  we would get 
 a contribution to $P_{12}$ represented in Fig.~9.
 However, this is equivalent to $P_{01}$ since the disconnected lines 
 just give a contribution proportional to $\delta (k-k')$.

 Similarly, it is easy to see that as long as we consider the emission or 
 absorption of at most a single photon we need to consider in ${\cal P}_{mn}$ 
 diagrams which are essentially equivalent to $P_{00}, P_{01}, P_{10},
 P_{11}^{(1)}$ multiplied by products of $\delta$-functions. 
 In this sense $P_{00}, P_{01}, P_{10}, P_{11}^{(1)}$
 constitute a minimal set of connected probability diagrams in terms of which
 all others involving a single photon can be constructed.
 We can sum all such contributions.
 Indeed,  consider an $S$-matrix element with $m$ soft photons in the initial 
 state and $n$ in the final state. 
 Suppose that $z$ of these are connected to the "hard" part.
 For example, in Fig.~8 $m=1, n=2$ and $ z=1$. 
 This is shown in Fig.~10, where the Green's function $F$ has disconnected 
 pieces. 
 Subdiagrams in which any of the $z$ lines are joined 
 together have been included in the shaded blob. 
 Thus, $z\le m+n$.
 
 Next consider $\sum_{i,f}|M_{mn}|^2$ . 
 We split  this up into a  sum of diagrams in which the connected and 
 disconnected pieces are  explicitly separated. 
 Note that this is just a rearrangement of the  perturbation series. 
 Thus, we first separate out the connected ones, i.e., 
 those for which all the photon lines are connected to the hard part. 
 We  are then left with the disconnected photons. 
 Let $G$ denote this disconnected Green's function with $a$ incoming and $b$ 
 outgoing photon lines.
 Then,  $\sum_{i,f}|M_{mn}|^2$ decomposes into a sum of terms  each of which 
 is a product of diagrams  of the type shown  in Figs.~11 and 12 and an 
 associated combinatoric factor. 
 Fig.~12 represents the connected part of ${\cal P}_{mn}$ which we denote 
 by $D_c(a ,b ,z,z')$. 
 Fig.~11 represents the associated disconnected part which will be denoted 
 by $D_{d}(m-a, n-b )$. 
 For this diagram the combinatoric factor is 
\begin{equation}
\left( \begin{array}{c} m \\ a \end{array} \right) a! 
\left( \begin{array}{c} n \\ b \end{array} \right) b! 
\label{combinatoric}
\end{equation}
 Note that the factor $a! b!$ is just the number of ways of connecting 
 all photon lines in a single curve that can cross back and forth 
 the unitarity cut. 
 Consequently,
\begin{eqnarray}
{\cal P}_{mn}&=&{1\over m!~ n!}\sum_{i,f}|M_{mn}|^2~ 
\nonumber \\
 &=& \sum_{a ,b} \frac{1}{(m-a)!~(n-b)!}
 D_{d}(m-a ,n- b )\sum_{z,z'}D_c(a ,b ,z,z') \, .
\label{lnprop}
\end{eqnarray}

 For up to one soft photon emission or absorption, it follows
 from our earlier discussions that we need to consider the following cases:
\begin{eqnarray}
&(1)&~~z~=~z'~=~0,~~~~D_c~=~P_{00}\nonumber \\
&(2)&~~z~=~z'~=~1,~b~=~a +1,~~~~D_c~=~P_{01}\\
&(3)&~~z~=~z'~=1,~a~=~b+1,~~~~D_c=P_{10}\nonumber \\
&(4)&~~z~=~0,2;~z~=~2,0;~~a~=~b,~~~~D_c~=~P_{11}^{(1)}\nonumber
\label{cases}
\end{eqnarray}
 Thus, putting together everything that is relevant to one photon 
 probabilities we have an expression for the LN probability 
 ${\cal P}_{mn}$ of the form
\begin{eqnarray}
{\cal P}_{mn}~=~{1\over m!~n!}D_{d}(m,n)P_{00}
              +\sum_{a}{1\over (m-a )!(n-a -1)!}D_{d}(m-a ,n-a-1) P_{01}
\nonumber \\
+ \sum_{a}{1\over (m-a -1)!(n-a )!} D_{d}(m-a -1,n-a )P_{10}\\
+\sum_{a}{1\over (m-a -1)!(n-a -1)!} D_{d}(m-a -1,n-a -1)P_{11}^{(1)}.
\nonumber
\label{adding}
\end{eqnarray}
Using
\begin{eqnarray}
{1\over m!~n!}D_{d}(m,n)~=~\sum_{a}{1\over (m-a )!~(n-a )!}D_{d}(m-a ,n-a )
\nonumber \\
~-~\sum_{a}{1\over (m-a -1)!~(n-a -1)!} D_{d}(m-a -1,n-a -1),
\label{teltric}
\end{eqnarray}
 we can rearrange the terms in Eq.~(\ref{adding}) to obtain
\begin{eqnarray}
{\cal P}_{mn}~=~\sum_{a}{1\over (m-a )!~(n-a )!} D_{d}(m-a ,n-a )P_{00}
\nonumber\\
+\sum_{a}{1\over (m-a )!~(n-a -1)!} D_{d}(m-a ,n-a -1)P_{01}
\nonumber  \\
+~\sum_{a}{1\over (m-a -1)!(n-a )!} D_{d}(m-a -1, n-a )P_{10} 
\nonumber \\
~+\sum_{a}{1\over (m-a -1)!~(n-a -1)!} D_{d}(m-a -1,n-a -1)
                                        (P_{11}^{(1)}-P_{00}).
\label{unite}
\end{eqnarray}
Hence, 
\begin{eqnarray}
{\cal P}~=~\sum_{m,n}{\cal P}_{mn}~=~\sum_{m,n}\sum_{a}\left(
{1\over (m-a )!~(n-a )!} D_{d}(m-a ,n-a )\right)
\nonumber \\ 
 \times \left( P_{00}+P_{01}+P_{10}+ (P_{11}^{(1)}-P_{00})\right)
\label{klnfac}
\end{eqnarray}
 The disconnected pieces have now factored out into a multiplicative factor.

 As mentioned above, the photon line in $P_{11}^{(1)}$ can be obtained 
 from that in $P_{00}$ by the replacement (\ref{repl}).
 Thus, in the combination $(P_{11}^{(1)}-P_{00})$, the photon propagator is
\begin{equation}
 2\pi\delta(k^2)-{i\over k^2+i\epsilon}~=~{-i\over k^2-i\epsilon}~=~
\left({i\over k^2+i\epsilon}\right)^*.
\label{cconj}
\end{equation}
 We can observe at this point the pattern of combining the $P_{ij}$
 to generate complex conjugate propagators. 
 The $P_{ij}$ that are added (subtracted) together correspond to diagrams 
 that differ in the number of cut photon loops. 
 By a cut photon loop we mean explicitly a photon line which begins and ends 
 on the same side of the unitarity cut after having crossed it at least once.
 We will call $(P_{11}^{(1)}-P_{00})$ to be $CP_{00}$. 
 What we have shown  is that the action of the operator $C$ on $P_{00}$ is 
 to replace the virtual photon propagator in $P_{00}$ by its 
 complex conjugate. 
 $P_{00}$ and $CP_{00}$ are in all other respects identical. 
 This then is the  complex conjugation rule alluded to earlier, i.e., 
 we may forget about the disconnected diagrams, consider only the virtual 
 and real emission diagrams and to these add the real absorption diagram and
 a virtual diagram in which the photon propagator has been replaced
 by its complex conjugate. 
 In the next section we will see that this simple result generalizes to 
 all orders.

\section{The KLN Factorization Theorem to All Orders}

 In this section, using algebraic methods, we will generalize the 
 discussion of the previous section to all orders in the 
 coupling and for an arbitrary process involving soft quanta.
 Even though we refer to photons explicitly below, the proof carries over for 
 other field theories as well, such as scalar QED or Yukawa theories with
 light isoscalars.
 The only restriction on the underlying field theory will be that 
 the soft particles have no inherent self interactions. 
 This latter case will be discussed in a separate publication \cite{asz1}. 

 Let ${\cal P}_{mn}$ denote the Lee-Nauenberg probability for the transition
 \begin{center} 
 $ A + m \ \mbox{soft~photons}\ \rightarrow 
   B + n \ \mbox{soft~ photons}\ $. 
 \end{center}
 In general,  and as seen explicitly in section 2,
 ${\cal P}_{mn}$ contains contributions coming 
 from disconnected diagrams.
 We denote by ${\cal P}^{(p)}_{mn}$ the probability obtained
 when up to $p$-photons are emitted or absorbed by the connected
 part of ${\cal P}_{mn}$ which contains the $A \rightarrow B$ transition.
 Furthermore, we define ${\cal P}^{(p)}$ as
 \begin{equation}
 {\cal P}^{(p)} = \sum_{m,n} {\cal P}^{(p)}_{m n} \, .
 \label{pdef}
 \end{equation}   
 Then the following theorem is true.

 Theorem:
 \begin{eqnarray}
 {\cal P}^{(p)} &=& \left\{ \sum_{m,n} \sum_{a} 
            \frac{1}{(m-a)! (n-a)!} \frac{(a+1)^{p-1}}{(p-1)!}
            D_d(m-a, n-a) \right\}  
 \nonumber \\
 &\times& \sum_{i=0}^{p} \sum_{j=0}^{p-i} \sum_{k=0}^{p-i-j}
           C^k P_{ij} \, ,
 \label{theorem}
 \end{eqnarray}
 where $D_d(m-a, n-a)$ is the disconnected part, $P_{ij}$ is 
 the connected part and the operator $C$ acts on $P_{ij}$ as
 follows:
 
 $C^0 P_{ij} = P_{ij}$

 $C^1$ acts on the cut diagram representing $P_{ij}$ by replacing it
 with an identical one, except that one virtual soft photon propagator
 has been replaced by its complex conjugate. The action of $C^1$ is
 repeated for all virtual photon propagators in $P_{ij}$. 

 $C^k$ replaces each and every subset of $k$ virtual soft photon propagators 
 in $P_{ij}$ by their complex conjugates.

 \noindent
 Note how the disconnected part has completely factorized from the 
 connected part. For given $p$, the connected part generates
 $(p^3/6 + p^2 + 11 p/6 +1)$ terms. They correspond to the points
 of a cubic lattice in $(ijk)$ space constrained by 
 $i, \, j, \, k\, \ge 0$ and $i+j+k \le p$.

 Proof: \\  
 For fixed $m$ and $n$, ${\cal P}^{(p)}_{mn}$ satisfies the 
 following topological decomposition
 \begin{eqnarray}
 {\cal P}^{(p)}_{mn} &=& 
 \frac{1}{m!\, n!} \,
 \sum_{a=0}^m  
 \left( \begin{array}{c} m \\ a \end{array} \right) \, a! 
 \sum_{b=0}^n 
 \left( \begin{array}{c} n \\ b \end{array} \right) \, b! \,
      D_d(m-a, n-b) 
 \nonumber \\
 &\times&
 \sum_{q=0}^p \, \sum_{z=0}^{2q} \, \sum_{z^\prime=0}^{2q-z} \,
 D_{c}(a, b; z, z^\prime) \, ,
 \label{topdec}
 \end{eqnarray}
 which has been already introduced in section 2 for the $p=1$ case.
 The structure of the disconnected part $D_d$ and the connected part
 $D_c$ are depicted in Figs.~11 and 12 respectively.
 Moreover, the $(a+b+z)$-point function $F$ to the left
 of the unitarity cut and the $(a+b+z^\prime)$-point
 function $F^\prime$ to the right are ${\cal O}(e^0)$, i.e.
 they contain no interactions. 
 The presence of fermion loops in the cut photon lines just lead 
 to multiplicative wave function renormalization factors.
 However, $F$ and $F^\prime$ do contain disconnected
 pieces of the form depicted in Fig.~13. 
 It is because of these disconnected pieces in the left and right
 amplitudes that, after squaring them, a photon line in $D_c$
 can cross the unitarity cut back and forth arbitrarily many times
 without changing the order of the interaction.
 Cutting a photon line of momentum $k^\mu$ many times results
 in one $(2 \pi) \delta(k^2)$ factor but the combinatorics of the
 disconnected piece are affected. 
 Consequently, the topological structure of $D_c(a, b; z, z^\prime)$
 is as depicted in Fig.~14. 
 The photon lines that cross the unitarity cut can be classified as
 follows. 
 There are lines belonging to $l$ cut loops whose both ends are
 connected to the left transition amplitude, lines belonging to
 $l^\prime$ cut loops connected to the right transition amplitude,
 and finally $r$ lines whose ends connect left and right amplitudes.
 With $q$ denoting the total number of cut photon lines 
 it is easy to see that the following relations hold.
 \begin{eqnarray}
 z+z^\prime=2 q, \; q=0,1,...p
 \nonumber \\
 z=2 l+r, \; z^\prime= 2 l^\prime + r
 \\
 l+l^\prime+r = q
 \nonumber
 \label{counting}
 \end{eqnarray}
 A configuration with given $l$, $l^\prime$ and $r$ gives contributions
 to the probability
 $ P_{l+l^\prime+i, l+l^\prime+r-i}, \, i=0,1,...r \, ,$
 because each cut loop yields one initial(absorbed) and one 
 final state (emitted) photon
 and the set of $r$-lines can give $i$-initial 
 and $(r-i)$-final state photons.
 The combinatoric factor in front of $D_d(m-a, n-b)$, Eq.~(\ref{topdec}),
 for fixed $q$, $r$ and $i$ becomes
 \begin{eqnarray}
 &\quad&
 \sum_{a_1=0}^{m} \,
 \sum_{a_2=0}^{m-a_1} \, ...
 \sum_{a_q=0}^{m-a_1...-a_{q-1}} \,
 \frac{1}{(m-a_1...-a_q-(q-r+i))!}
 \frac{1}{(n-a_1...-a_q-(q-i))!}
 \nonumber \\
 &=&
 \sum_{a=0}^m \frac{(a+1)^{q-1}}{(q-1)!}
 \frac{1}{(m-a-(q-r+i))!} \frac{1}{(n-a-(q-i))!}
 \label{combin}
 \end{eqnarray}
 where $a=\sum_{s=1}^q \, a_s$.
 There are $q$ sums in the above expression because there are
 $q$ photon lines, each one being taken back and forth across 
 the unitarity cut $a_s$ times.

 With $L=l+l^\prime$, the Lee-Nauenberg probability 
 ${\cal P}_{mn}^{(p)}$ in Eq.~(\ref{topdec}) can be written as
 \begin{equation}
 {\cal P}_{mn}^{(p)} = 
 \sum_{a=0} \,
 \sum_{r=0}^{p} \, 
 \sum_{i=0}^{r} \,
 \sum_{L=0}^{p-r} \, 
 \frac{(a+1)^{L+r-1}}{(L+r-1)!} \,
 \frac{D_d(m-a-L-i, n-a-L-r+i)}{(m-a-(L+i))! \, (n-a-(L+r-i))!} \,
  P_{L+i, L+r-i} \, .
 \label{LNP}
 \end{equation}
 The connected probabilities depend only on $L$ because
 we do not distinguish whether the cut loops are attached to
 the left or the right transition amplitude.
 In order to prove theorem (\ref{theorem}) we must combine
 the terms in Eq.~(\ref{LNP}) in a specific manner.
 To this end we will use the following telescopic relation
 for the series in $a$.
 \begin{eqnarray}
 &\quad&
 \sum_a \, \frac{(a+1)^{L+r-1}}{(L+r-1)!}
 \frac{D_d(m-a-L-i, n-a-L-r+i)}{(m-a-L-i)! (n-a-L-r+i)!}
 \nonumber \\
 &=&
 \sum_a \, \frac{(a+1)^{L+r}}{(L+r)!}
 \left[
 \frac{D_d(m-a-L-i, n-a-L-r+i)}{(m-a-L-i)! (n-a-L-r+i)!} \right.
 \nonumber \\
 &\quad&
 \hspace{2.5cm} - \left.
 \frac{D_d(m-a-L-i-1, n-a-L-r+i-1)}{(m-a-L-i-1)! (n-a-L-r+i-1)!}
 \right]
 \label{tric}
 \end{eqnarray}
 This is a generalization of the expression in Eq.~(\ref{teltric}).
 Let us now consider the contribution to ${\cal P}^{(p)}_{mn}$
 for fixed $r$ and $i$ in Eq.~(\ref{LNP}) and apply the above
 relation $p-r-L$ times. 
 The result is
 \begin{eqnarray}
 \left. {\cal P}_{mn}^{(p)} \right|_{ r, i: {\rm ~fixed}}
 &=&
 \sum_{L=0}^{p-r} \, \sum_{f=0}^{p-r-L} (-1)^f
 \left( \begin{array}{c} p-r-L \\ f \end{array} \right)
 \sum_{a}
 \frac{(a+1)^{p-1}}{(p-1)!} 
 \nonumber \\
 &\times&
 \frac{D_d(m-a-L-i-f, n-a-L-r+i-f)}
      {(m-a-(L+i+f))! (n-a -(L+r-i+f))!} \, 
  P_{L+i, L+r-i}
 \label{manytrics}
 \end{eqnarray}
 Then, we group together terms with the same  $L+f=k$ in the above expression.
 This yields
 \begin{equation}
 \left. {\cal P}_{mn}^{(p)} \right|_{ r, i: {\rm ~fixed}} =
 \sum_{a} \frac{(a+1)^{p-1}}{(p-1)!} \, 
 \sum_{k=0}^{p-r} \,
 \frac{D_d(m-a-k-i, n-a-k-r+i)}
      {(m-a-k-i)! (n-a-k-r+i)!} \,
 C^k P_{i, r-i}
 \label{partsum}
 \end{equation}
 where we have identified
 \begin{equation}
 C^{k} P_{i, r-i} = \sum_{f=0}^{k} (-1)^f 
 \left( \begin{array}{c} p-r-k+f \\ f \end{array} \right)
 P_{i+k-f, r-i+k-f}
 \label{Cdef}
 \end{equation}
 To obtain ${\cal P}_{mn}^{(p)}$ we sum the expression in 
 Eq.~(\ref{partsum}) over $r$ and $i$ and
 to obtain the probability ${\cal P}^{(p)}$ defined in Eq.~(\ref{pdef})
 we sum also over $m$ and $n$.
 After appropriate shifts in the variables $m$, $n$ the net result is
 \begin{eqnarray}
 {\cal P}^{(p)} &=&
  \left\{ \sum_{m,n} \sum_{a} 
  \frac{1}{(m-a)! (n-a)!} \frac{(a+1)^{p-1}}{(p-1)!}
            D_d(m-a, n-a) \right\}  
 \nonumber \\
 &\times&
 \sum_{r=0}^p \, \sum_{i=0}^{r} \, \sum_{k=0}^{p-r} C^k P_{i, r-i}
 \label{proof}
 \end{eqnarray}
 Upon change of variables $(i, r) \rightarrow (i, j=r-i)$ equation
 (\ref{theorem}) is retrieved.

 Finally we must show that the operator $C$ whose action was
 defined in Eq.~(\ref{Cdef}) acts as described in the theorem. 
 Obviously, $C^0 P_{i, r-i} = P_{i, r-i}$.
 For $k=1$, Eq.~(\ref{Cdef}) gives
 \begin{equation}
 C P_{i, r-i} = P_{i+1, r-i+1} - (p-r) P_{i, r-i} \, .
 \label{k=1}
 \end{equation}
 We observe that $P_{i+1, r-i+1}$ and $P_{i, r-i}$ have the same
 number of $r$-lines with $i$ of them being incoming (absorbed).
 They differ in that $P_{i+1, r-i+1}$ has in addition one cut
 photon loop.
 This can occur when one of the remaining $(p-r)$ virtual soft
 photon propagators of $P_{i, r-i}$ has crossed the unitarity cut
 and turned into a $\delta$-function in $P_{i+1, r-i+1}$.
 When the two probabilities are added as in Eq.~(\ref{k=1}) the
 result is that each soft photon propagator of momentum $k^\mu$ in
 $P_{i, r-i}$ is substituted by its complex conjugate according
 to the relation
 \begin{equation}
 (2 \pi) \delta(k^2) - \frac{i}{k^2+i\epsilon} = -\frac{i}{k^2-i\epsilon} \, .
 \label{cc}
 \end{equation} 
 The combinatoric factor $(p-r)$ in front of $P_{i, r-i}$ indicates
 that $CP_{i, r-i}$ generates $(p-r)$ terms, each term being equal to
 $P_{i, r-i}$ with one of its soft propagators complex conjugated.
 There are actually $(p-r)$ terms also in $P_{i+1, r-i+1}$ differing
 in the way the cut loop connects to the rest of the graph, but
 we have lumped them all together in $P_{i+1, r-i+1}$.
 For $p=1$, the definition (\ref{Cdef}) for  $C P_{00}$ generates the
 two terms $P_{11}^{(1)}$ and $-P_{00}$ already encounterd in section 2.

 For arbitrary $k$ the action of $C^k$ can be established in a similar way as
 above.
 The sum over $f$ in Eq.~(\ref{Cdef}) contains all connected probabilies
 starting with $ P_{i+k, r-i+k}$ with $k$ cut loops and ending with
 $P_{i, r-i}$ whose soft photon lines are all virtual except the $r$ 
 cut lines connecting left and right subdiagrams, 
 common to all $P$'s in the sum.
 Let us call $k_1^\mu, ... k_k^\mu$ the momenta of photons in the
 cut loops.
 Then Eq.~(\ref{Cdef}) generates the following momentum factor
 \begin{equation}
 \sum_{f=0}^k (-1)^f
 \prod_{l=1}^{k-f} (2 \pi) \delta(k_l^2)
 \prod_{s=1}^{f} \frac{i}{k^2_s+i\epsilon} 
 =
 \prod_{l=1}^k \left(\frac{i}{k^2_l+i \epsilon} \right)^*
 \label{conjugate}
 \end{equation}
 where the sets $\{k_l^\mu \}$ and $\{k_s^\mu \}$ on the left-hand side 
 are disjoint.
 Just like in the $k=1$ case, we have not included in the above expression
 the combinatoric factor in front of every $P_{i+k-f, r-i+k-f}$.
 This factor counts the inequivalent ways of identifying the momenta
 of the virtual propagators in each $P$ with the set of momenta
 $\{k_s^\mu\}$.
 The combinatoric factor in front of $P_{i, r-i}$ is 
 \begin{equation}
 \left( \begin{array}{c}  p-r \\ k \end{array} \right)
 \end{equation}
 which indicates that all different $k$-plets of soft virtual photon 
 propagators are substituted by their complex conjugates.
 This completes the proof of the theorem.

 Thus we conclude that, even to all orders, a very simple prescription
 emerges on how to account for the KLN sum including the effect
 of the disconnected diagrams in covariant perturbation theory. 
 We first write down all the cut diagrams contributing
 at the level of the inclusive cross section. 
 Next we include diagrams in which each (soft) emission external line 
 is changed to a (soft) absorption line, and in addition each soft propagator
 is successively changed to its complex conjugate. 
 
 Applying the theorem for the case of two photons we obtain:
\begin{eqnarray}
{\cal P}^{(2)}= \left( \sum_{m,n}\sum_{a}{1 \over (m-a)!(n-a)!}
(a+1)D_d(m-a,n-a) \right) \cdot 
\nonumber \\
\left(P_{00}+P_{01}+P_{02}+P_{20}+P_{11}+CP_{00}+CP_{01}+CP_{10}+C^2 P_{00} 
\right),
\label{twophotons}
\end{eqnarray}
 which reproduces the result given in Ref.~\cite{az3}.

\section{ The Low Theorem for Multi-Photon Processes }

 Low's theorem  states that the leading and the next to leading
 terms of order $1/{\omega}$  and $\omega^0$ respectively, 
 $\omega$ being the photon energy, in a general bremsstrahlung 
 amplitude can be expressed in terms of the corresponding
 radiationless amplitude. 
 It applies in a straightforward manner to spinless charged particles, 
 and has been extended to the case of spin-$1/2$ particles in \cite{kb},
 where it was shown that in the low energy limit the  ${\cal O}(\omega^0)$  
 structure dependent contributions present in the amplitude
 do not contribute to the unpolarized cross sections. 
 This is known as the Burnett-Kroll theorem.
 The Low theorem has been extended to multi-photon (emission or absorption)
 processes in \cite{bg,dd} by utilizing the fact that in an abelian theory the
 Ward identities act independently on different photons.
 More recently some issues have been raised concerning the range of validity 
 of the Low theorem in the high energy limit. 
 Suppose that the quarks have a mass $m$ and we are interested in 
 soft photons with energy $\omega < m$. It was shown in \cite{dd}
 that though in the high energy limit Low's theorem is valid in the domain
 $0 \leq \omega \leq m^2/Q$,  it may be extended to the larger region 
 $0 \leq \omega \leq m$. 
 It is the next to leading term in the Low and the Burnett-Kroll theorems 
 which becomes sensitive to the structure of the jets of the external 
 charged particles. 
 For these subtleties and for a modern treatment of the Low theorem
 the reader is referred to Ref.~\cite{dd}.

 Our interest in the Low theorem arises in its application to the study of the
 leading power corrections to various processes, which are known to arise 
 due to soft radiation. 
 To avoid unnescessary complications due to collinear divergences we will 
 give the quarks a small mass $m$ and consider soft radiation of 
 energy $\omega < m$.  
 The Low theorem allows us to express terms of order $1/{\omega^2}$ 
 and $1/{\omega}$ in the matrix element squared in terms of the square of the 
 radiationless amplitude. 
 By power counting we can isolate not only the logarithmic but the 
 linear dependence, if any, on an infrared cutoff as well. 
 The Low theorem thus turns out to be very useful for studying the leading 
 $1/Q$ power corrections in gauge theories \cite{az3}.
 In this section, for completeness, we will outline the construction of the
 photon amplitude in terms of the radiationless one up to the next to
 leading order. 

 In the following we will denote the radiationless (elastic) amplitude for
 the process $q+\bar{q} \rightarrow \gamma^*$ by $M_{el}$.
 In the high energy limit the radiationless amplitude can be written in 
 a factorized form which is a product of the external jet factors, the hard 
 amplitude and the soft subdiagrams.
 For explicit definitions of these factors see Refs.~\cite{sud}.
 We will include the soft subdiagrams inside the hard amplitude.
 For simplicity of presentation we will ignore the jet structure of
 the external lines and consider the single particle approximation for them. 
 Explicit expressions for the amplitude for single photon emission 
 to be given below will also involve the elastic vertex function
 $\Gamma_{el}$, related to $M_{el}$ by
\begin{equation}
M_{el}= \bar{v} \Gamma_{el} u \, ,
\label{spinors}
\end{equation}
 where $\bar{v}$ and $u$ are the external spinors, which, for convenience
 of notation, are assumed to absorb any wave function renormalization factors. 
 Then the amplitude for the emission of one photon with momentum $k$ and
 polarization $\epsilon^\mu$ can be written as 
\begin{equation}
M_{01,\mu}\epsilon^{\mu} ~=~ M_{\mu}^{ext}\epsilon^{\mu} +
M_{\mu}^{H}\epsilon^{\mu} \, ,
\label{ampl}
\end{equation}
 where the first term is the amplitude for the emission of the photon from the
 external line and the second one from the hard subdiagram.
 Through the use of the abelian Ward identity 
 and charge conservation we can write the amplitude for the emission
 of a single soft photon up to terms of order $\omega^0$ as 
\begin{eqnarray}
M_{01,\mu} \epsilon^{\mu} = 
\bar{v} \epsilon^{\mu} 
\left[
\sum_{a=1,2}Q_a\left({ {(2p_a-k)_{\mu}} \over {2p_a\cdot k }}
- G^{\lambda}_{\, \mu}(p_a,k) {\partial \over \partial p_a^{\lambda}}
\right) \Gamma_{el} \right] u
\nonumber \\
+\bar{v} \Gamma_{el} \left[Q_1 + ( \not \! p_1 +m_1){ \kappa_1 \over 2m_1}
                    \right]
{{ \not k \not \epsilon } \over {2 k\cdot p_1}}u  
\nonumber \\
+\bar{v}{{ \not k \not \epsilon } \over {2k\cdot p_2}}
 \left[Q_2 + (\not \! p_2 + m_2) { \kappa_2 \over 2 m_2}\right] 
\Gamma_{el}u \, . 
\label{low}
\end{eqnarray}
 In the above, $p_a$ and $\kappa_a$, $a=1,2$ are the momenta and the 
 anomalous magnetic moments  of the incoming fermions, 
 $Q_a$ are their charges with $\sum Q_a=0$, and
\begin{equation}
G^{\lambda}_{\, \mu}(p_a,k)~=~g^{\lambda}_{\, \mu}
  -{ k^{\lambda}(2 p_a-k)_{\mu} \over {2 p_a\cdot k}} \, . 
\label{Gpol}
\end{equation}
 The last two terms in Eq.~(\ref{low}) are ${\cal O}(\omega^0)$ and depend
 on the spin of the external particles. 
 They do not contribute to the corresponding 
 $P_{01}$ in accordance to the Burnett-Kroll theorem. 

 For the case of multiple photon emission, as a consequence of the fact that 
 the abelian Ward identity acts independently on each photon,  one can show 
 \cite{dd} that the amplitude including the next to
 leading terms may be written as 
\begin{eqnarray}
M_{0n,[\mu_1 \mu_2...\mu_n]} \prod_{i=1}^n\epsilon^{\mu_{i}}(k_i) &=&
\bar{v} \prod_{i=1}^{n} \epsilon^{\mu_i} 
\left[
\sum_{a_i=1,2}Q_{a_i}
\left({ {(2p_{a_i}-k_i)_{\mu_i}} \over {2p_{a_i}\cdot k_i }}
- G^{\lambda}_{\, \mu_i}(p_{a_i},k_i) {\partial 
                                        \over \partial p_{a_i}^{\lambda}}
\right) \Gamma_{el} \right] u 
\nonumber \\
&+& ({\rm jet~structure~dependent~terms}) \, .
\label{multiphoton}
\end{eqnarray}
 The square brackets on the left hand side denote symmetrization. 
 The jet structure dependent terms 
 can also be written as a product of independent emissions from
 the external fermion lines. 
 This  formula, which is essentially one of independent emission for the 
 next to leading terms as well, has important consequences for 
 the $1/Q$ corrections as we discuss in the next section.

\section{Fate of the $1/Q$  Terms in the Drell-Yan Process}

 In this section we will apply the results of the previous sections to 
 investigate if there are any $1/Q$ power corrections to 
 the inclusive Drell-Yan process. 
 We will begin with the case of single photon processes.
 In order to make the connection to the previous sections we note that from 
 Eq.~(\ref{dy}) the hard inclusive cross section at the 
 parton level, i.e., $W_{q\bar{q}}$ may essentially be identified with the sum 
 $P_{00}+P_{01}$ for the process 
 $q + \bar{q} \rightarrow \gamma^* + {\rm photons} $.
 This is because only the emission and virtual diagrams
 (and not the absorption process) contribute to a physical cross section.

 From the KLN theorem, however, we know in particular that the sum
 $(P_{00}+P_{01}+P_{10}+CP_{00})$ does not contain terms logarithmic 
 or linear in an infrared cutoff $\lambda$ \cite{asz}.
 From this we must show the absense of any logarithmic or linear dependence on
 the  infrared cutoff in $W_{q\bar{q}} $. 
 This we now proceed to do using the results of the previous section on the 
 Low theorem. 
 The cancellation of terms which are proportional to $\ln \lambda$ is in fact 
 nothing but the Bloch-Nordsieck mechanism. 
 To see how it follows from the KLN theorem we observe the following. 
 First, from power counting, it is easy to see that to logarithmic
 accuracy, $P_{01} \approx P_{10}$.  
 Next we note that to find the logarithmically divergent terms in the virtual 
 diagrams, we may replace the virtual photon propagator by 
 $2 \pi \delta(k^2)$. 
 Since the only difference between $P_{00}$ and $CP_{00}$ is that the 
 virtual photon propagator in the first is complex conjugated in the second, 
 it follows that again up to logarithmic accuracy, $P_{00} \approx CP_{00}$ . 
 Therefore, the absense of any $\ln \lambda$ terms in the sum 
 $(P_{00}+P_{01}+P_{10}+CP_{00})$ implies its absense in the physical
 inclusive cross section $(P_{00}+P_{01})$. 

 Next we show the absense of any terms proportional to $\lambda$ in this 
 physical cross section.  
 For this we note that it is known that any possible linear terms
 can only arise from the real emission diagrams \cite{cs,bb}.  
 Thus we must now show  that  up to finite terms  as $\lambda \rightarrow 0$ 
 and up to terms of order $\lambda^2$, $P_{01}$ and $P_{10}$ are equal.

 To see this let us consider the lowest order emission amplitude $M_{01}$.
 This can be expressed up to terms $\omega^0$ as
\begin{equation}
M_{01, \mu} ~=~e{(2p_2 -k)_{\mu}\over 2p_2\cdot k- k^2 } M_{el}
 - e{(2p_1-k)_{\mu}\over 2p_1\cdot k-k^2} M_{el}
  +e(p_1+p_2)\cdot k
\left( {p_{1\mu}\over p_1\cdot k}-{p_{2 \mu}\over p_2\cdot k}\right)
{\partial M_{el} \over \partial (p_1 \cdot p_2)} \, ,
\label{1photon}
\end{equation}
 where we have set $Q_q=-Q_{\bar{q}}=e$. 
 In the above expression we have kept only 
 the relevant terms that contribute to the emission probability
 up to ${\cal O}(1/ \omega)$.
 The jet structure dependent terms, after mass, charge and wave function 
 renormalization can be shown not to contribute to the emission 
 probability terms that are linear in the IR cutoff $\lambda$. 
 Next, in calculating $\sum |M_{01}|^2$ we keep terms $O(1/k^2)$ and 
 $O(1/k)$ and in this way arrive at the expression 
\begin{equation}
 P_{01}~=~
{\alpha \over \pi}|M_{el}|^2
\int{ {\rm d}^3 k\over 2 \omega}
\delta \left( (p_1+p_2-k)^2-Q^2\right)
 {2Q^2\over 2(p_1\cdot k)2(p_2\cdot k)} \, \Theta(k^2_\perp -\lambda^2) \, ,
\label{onephoton}
\end{equation}
 This agrees with the one loop expression given in \cite{bb}.
 A similar expression can be obtained for $P_{10}$ and we find here
\begin{equation}
P_{10}~=~{\alpha \over \pi}|M_{el}|^2
\int{ {\rm d}^3k\over 2 \omega} \delta\left((p_1+p_2+k)^2-Q^2\right)
{2Q^2\over 2 (p_1\cdot k)  2 (p_2\cdot k)} \, \Theta(k^2_\perp -\lambda^2) \, .
\label{p01}
\end{equation}
 Comparing the two probabilities upon taking moments with respect to $x$,
 Eq.~(\ref{dy}), we obtain
\begin{equation}
 P_{01} = P_{10} + {\cal O}(\lambda^2) + {\cal O}(\lambda^0) 
 {\rm ~regular~terms} \, .
\label{compare}
\end{equation}
 Then it follows from the KLN theorem that $P_{01}$, which is relevant for
 the DY cross section, has no terms linear in infrared cut off $\lambda$.
 This was shown in \cite{bb} by explicitly analyzing the integral in
 Eq.~(\ref{onephoton}). 
 Here we note that this absense is a direct consequence of the KLN 
 and the Low theorems.

 In fact in the abelian theory it is easy to see that the absense of any terms
 linear in $\lambda$ at the level of one photon emission also implies 
 its absense to all orders in the perturbation expansion. 
 This is because for multi-photon emission we saw in the previous section 
 that the amplitude up to the next to leading terms is essentially a product 
 of those for the individual photons, see Eq.~(\ref{multiphoton}).
 Then, $1/Q$ corrections relative to the leading term can only arise
 by a single photon emission at a time in the above product.
 But, as we saw earlier, such an emission does not generate $1/Q$ corrections.
 We thus conclude that at least in the abelian theory there are no $1/Q$ 
 corrections to the Drell-Yan process.

 \section{Conclusions} 

 In this paper we have developed a novel formulation of the KLN theorem for 
 the case of soft radiation. 
 The purely disconnected contributions to transition 
 probabilities can be factored out from the connected ones. 
 The effect of the disconnected diagrams to the connected probabilities 
 can be encoded into a simple rule involving just the emission diagrams and 
 the virtual ones together with their radiative corrections. 
 Then this was used in conjunction with the Low theorem to show how the 
 terms linear in an infrared cutoff are cancelled  in inclusive processes. 
 For the latter, the case of the Drell-Yan process was discussed as an 
 example. 
 We would like to emphasize that the crucial point of the derivation above can
 be generalized to other cases as well.
 First, the KLN theorem  applies not only to logarithmic divergences but
 also to the leading power suppressed terms and the KLN factorization
 theorem referred to above was shown to hold for an arbitrary procees to
 all orders. 
 Second, linear terms arise because of the contribution of soft,
 not collinear photons. 
 This means that the amplitudes of emission and absorption refer to energies 
 that are close to each other. 
 Although one cannot simply neglect the energy difference as in case of 
 logarithmic divergences, one can apply the Low theorem to relate the 
 emission and absorption amplitudes to the desired accuracy. 
 As a result, the KLN theorem turns into a statement of cancellation upon 
 summation over final states alone, similar to the Bloch-Nordsieck case 
 for any arbitrary process.
 We should point out, however, that there are some subtleties in the
 application of the Low theorem to high energy processes. 
 These have to do with the fact that the next to leading contributions to the 
 Low theorem at high energy become sensitive to the jet structure of the 
 external lines.
 These questions have been extensively discussed in \cite{dd} where it was
 shown that in the abelian theory one obtains the formula for independent
 emission (or absorption) even upon extending the Low theorem to high
 energy processes.  

 In conclusion we would like to emphasize that in this paper only the
 case of abelian field theories was discussed. 
 The non-abelian theory brings in new features which will be discussed in a
 separate paper.  

\section{Acknowledgements} 

 We would like to thank  A. H. Mueller and G. Sterman for many useful 
 discussions. 
 This work was supported in part by the US department of energy.

\newpage
\begin{center}
{\bf Figure Captions}
\end{center}

\begin{itemize}

\item
 Fig.~1. Radiationless amplitude for $q+\bar{q} \rightarrow \gamma^*$.

\item
 Fig.~2. Cut diagram representing the probability $P_{00}$ for 
 $q + \bar{q} \rightarrow \gamma^*$.

\item
 Fig.~3. The amplitude for $q+\bar{q} \rightarrow \gamma^* +
 {\rm soft~photon}$. 

\item
 Fig.~4. Cut diagram for the probability $P_{01}$.

\item 
 Fig.~5. Cut diagram for the probability $P_{10}$.

\item
 Fig.~6. Connected (a) and disconnected (b) diagrams for the amplitude \\
   $q+\bar{q}+ {\rm soft~photon} \rightarrow \gamma^* 
  + {\rm soft~photon}$.

\item
 Fig.~7. Cut diagrams representing the contributions to $P_{11}^{(1)}$.

\item
 Fig.~8. Disconnected diagram for the amplitude  \\
 $ q+\bar{q}+ {\rm soft~photon} \rightarrow \gamma^* 
                                      + {\rm two~soft~photons}$.   
\item
 Fig.~9. Cut diagram where the soft photon line crosses the unitarity cut 
 twice. This contribution is equivalent to $P_{01}$.

\item
 Fig.~10. Topological decomposition of the amplitude \\
 $ q+\bar{q}+{\rm m~soft~photons} \rightarrow \gamma^*
  +{\rm n~soft~photons}$.

\item
 Fig.~11. The disconnected part of the Lee-Nauenberg probability 
      ${\cal P}_{mn}$.

\item
 Fig.~12. The connected part of the Lee-Nauenberg probability 
      ${\cal P}_{mn}$. In the Drell-Yan case we identify $A$ with
      $q \bar{q}$ and $B$ with the decay products of $\gamma^*$.

\item
 Fig.~13. The structure of the $(a+b+z)$-point function $F$.

\item
 Fig.~14. The structure of the connected part $D_c(a,b;z,z^\prime)$.

\end{itemize}

\newpage

\centerline{
\epsfxsize=0.45\textwidth \epsffile{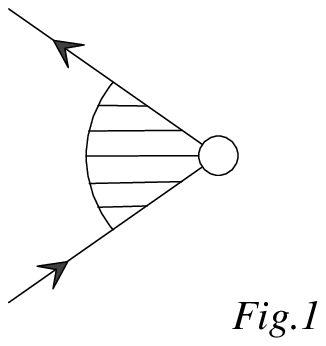} 
\hspace{0.1\textwidth} 
\epsfxsize=0.45\textwidth \epsffile{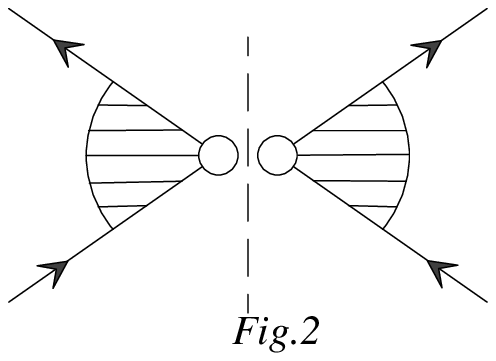} 
} 

\vspace{1.0cm} 

\centerline{
\epsfxsize=0.45\textwidth \epsffile{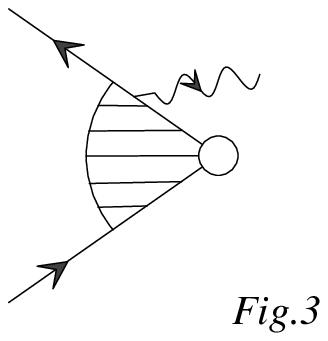} 
\hspace{0.1\textwidth} 
\epsfxsize=0.45\textwidth \epsffile{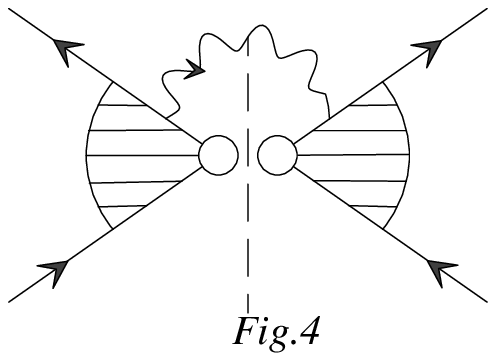} 
} 

\vspace{1.0cm} 

\centerline{
\epsfxsize=0.45\textwidth \epsffile{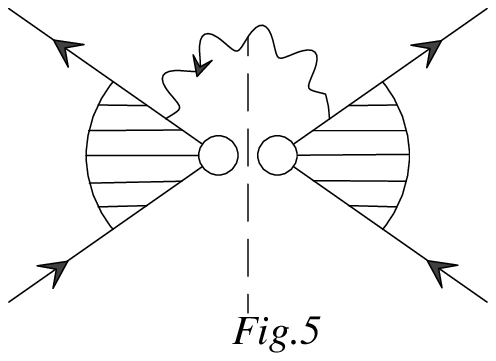} 
} 

\newpage

\centerline{
\epsfxsize=0.45\textwidth \epsffile{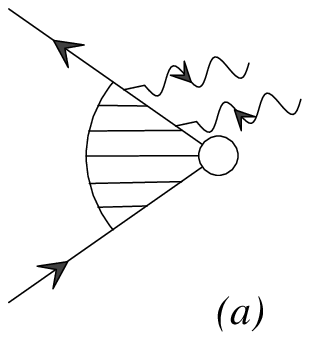} 
\hspace{0.1\textwidth} 
\epsfxsize=0.45\textwidth \epsffile{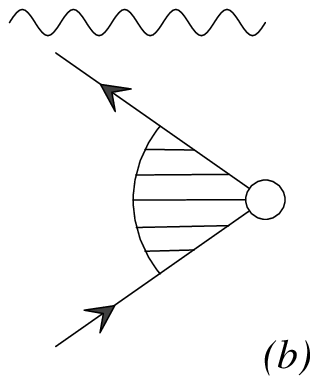} 
} 
\begin{center}
{\large \bf Fig.~6}
\end{center}

\vspace{1.0cm} 

\centerline{
\epsfxsize=0.45\textwidth \epsffile{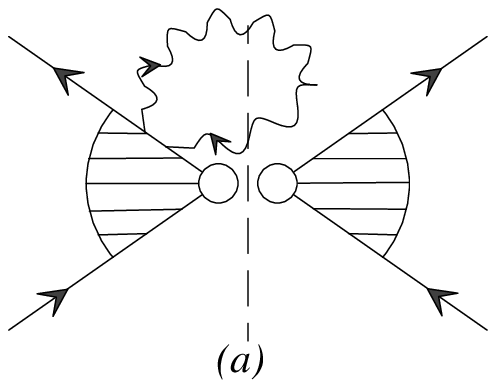} 
\hspace{0.1\textwidth} 
\epsfxsize=0.45\textwidth \epsffile{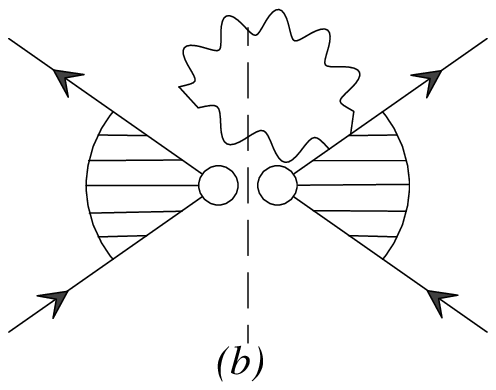} 
} 
\begin{center}
{\large \bf Fig.~7}
\end{center}

\vspace{1.0cm} 

\centerline{
\epsfxsize=0.45\textwidth \epsffile{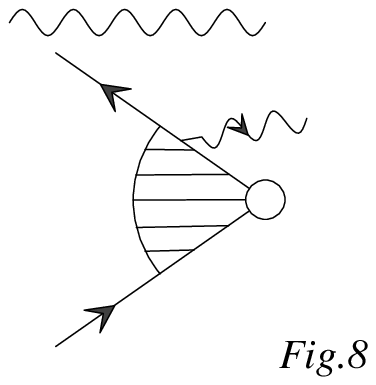} 
\hspace{0.1\textwidth} 
\epsfxsize=0.45\textwidth \epsffile{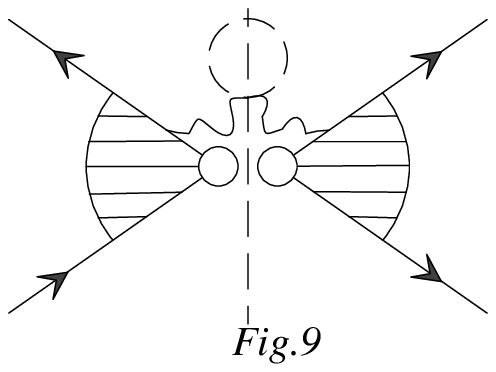} 
} 

\newpage 

\centerline{
\epsfxsize=0.45\textwidth \epsffile{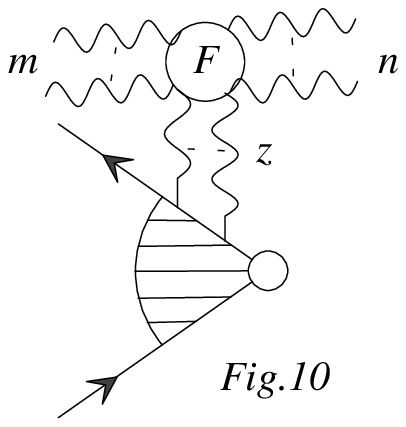} 
\hspace{0.1\textwidth} 
\epsfxsize=0.45\textwidth \epsffile{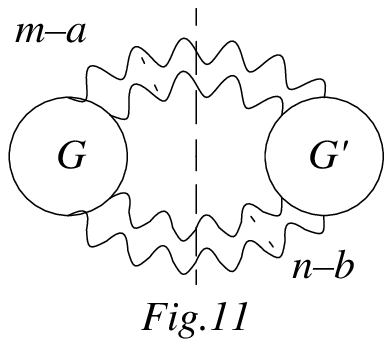} 
} 

\vspace{1.0cm} 

\centerline{
\epsfxsize=0.45\textwidth \epsffile{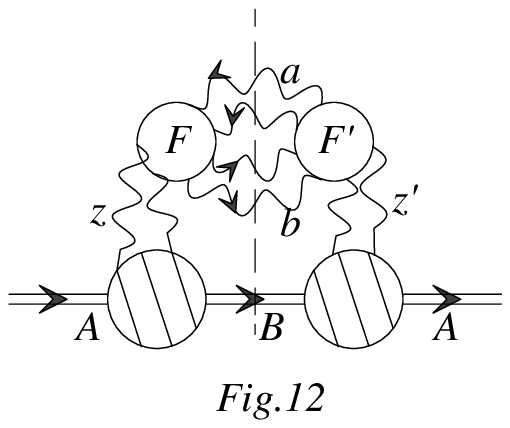} 
\hspace{0.1\textwidth} 
\epsfxsize=0.45\textwidth \epsffile{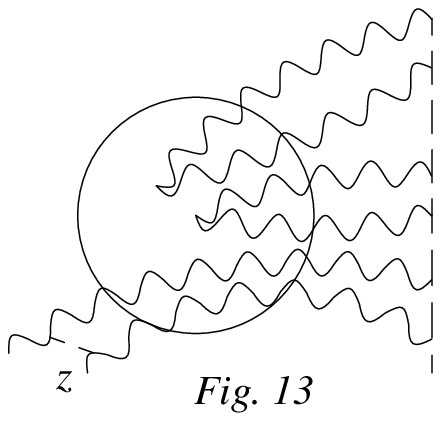} 
} 

\vspace{1.0cm} 

\centerline{
\epsfxsize=0.45\textwidth \epsffile{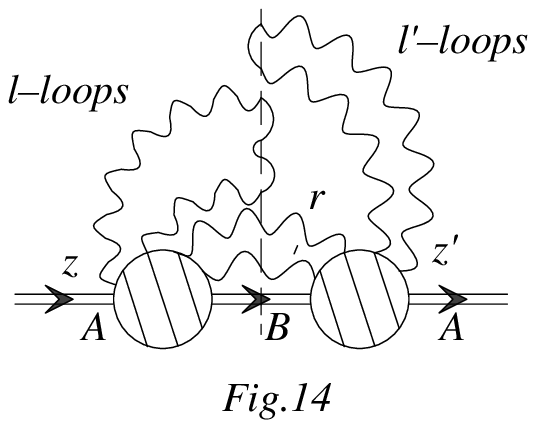} 
}

\end{document}